\documentclass[10pt]{article}
\usepackage[T2A]{fontenc}
\usepackage[cp1251]{inputenc}
\usepackage[english]{babel}
\usepackage{amssymb}
\usepackage{amsmath}
\usepackage{amsthm}
\usepackage{amsfonts}
\usepackage{amssymb}

\newtheorem{theorem}{Theorem}

\newtheorem{lemma}{Lemma}

\newtheorem{proposition}{Proposition}

\newtheorem{definition}{Definition}

\newtheorem{example}{Example}
\newtheorem{remark}{Remark}

\begin{document}
\begin{center}
\textbf{Stochastic L\'evy Differential  Operators and Yang-Mills Equations}\\
Boris O. Volkov\\

Steklov Mathematical Institute of Russian Academy of Sciences,\\ ul. Gubkina 8, Moscow, 119991 Russia\\E-mail: borisvolkov1986@gmail.com

\end{center}

{\bf Abstract:} The relationship between the Yang-Mills equations and the stochastic analogue of L\'evy differential operators is studied. The value of the stochastic L\'evy Laplacian is found by means of C\`esaro averaging of directional derivatives on the stochastic parallel transport. It is shown that the Yang-Mills equations and the L\'evy-Laplace equation for such Laplacian
are not equivalent as in the deterministic case. An equation equivalent to the Yang-Mills equations is obtained. The equation contains the stochastic L\'evy divergence. 
 It is proved that the Yang-Mills action functional can be represented as an infinite-dimensional analogue of the Direchlet functional of chiral field. This analogue is also derived using C\`esaro averaging.

keywords: L\'evy Laplacian, L\'evy divergence, stochastic parallel transport, Yang-Mills equations

AMS Subject Classification: 70S15, 81S25

\section*{Introduction}

The original L\'evy                                                                                                                                                                                                                                                                                                                                                                                                                                                                                                                                                                                                                                                                                                                                                                  Laplacian was defined for functions on the space  $L_2(0,1)$ in the twenties of the last century.
Paul L\'evy suggested different approaches to define this operator. The value of the L\'evy Laplacian 
on a function can be determined as an integral functional generated by the special form  of the second-order
derivative or as the C\`esaro mean of second directional derivatives along  vectors of some orthonormal basis
in $L_2(0,1)$ (see~\cite{L1951,F2005,Polishchuk}). Both approaches are used to define Laplacians acting on functions on the various functional spaces and these differential  operators  are also called L\'evy Laplacians.
 One of them has been defined as an integral functional in~\cite{AGV1993,AGV1994} by L.~Accardi, P.~Gibilisco, I.V. Volovich  and has been used for the description of  solutions of the Yang-Mills equations.  In these papers it has been proved that the connection in the trivial bundle over a Euclidean space satisfies the Yang-Mills equations if and only if the corresponding parallel transport is a harmonic functional for the L\'evy Laplacian (see also~\cite{Accardi}). The case of a nontrivial bundle over a compact Riemannian manifold has been considered in~\cite{LV2001} by R. Leandre and I. V. Volovich.
Also in~\cite{LV2001} the stochastic L\'evy Laplacian has been defined as a special integral functional.
For this operator the similar theorem holds: the stochastic parallel transport satisfies the L\'evy-Laplace equation if and only if 
the corresponding connection satisfies the Yang-Mills equation.
 The present paper focuses on the second approach to the L\'evy Laplacian. We  study the stochastic 
 L\'evy Laplacian defined as the C\`esaro mean of the second directional derivatives, associated with this Laplacian divergence and their connection to the Yang-Mills equations. In the deterministic case 
 the second approach
 to the definition of the L\'evy Laplacian has been used for study  the Yang-Mills equations  in~\cite{Volkov2012,Volkov2015,VolkovD}.

Let us recall  the general scheme of the definition of  homogeneous linear differential
operators  from~\cite{ASF}.
Let $X,Y,Z$ be real normed vector spaces. 
Let $C^n(X,Y)$ be the space of $n$ times Freshet differentiable   $Y$-valued 
functions on $X$.
Then for every $x\in X$ it is valid  that $f^{(n)}(x)\in L_n(X,Y))$.\footnote{Everywhere below 
$L(X,Y)$  is  space of linear continuous mappings from $X$ to $Y$. The space  $L_n(X,Y)$ is defined by induction:
$L_1(X,Y)=L(X,Y)$ and $L_n(X,Y)=L(X,L_{n-1}(X,Y))$}
Let  $S$ be a linear mapping from $dom S\subset L_n(X,Y)$ to $Z$.
The domain $dom D^{n,S}$  of a differential operator   $D^{n,S}$ of order $n$ generated by the operator $S$
consists of all $f\in C^n(X,Y)$ such that $f^{(n)}(x)\in dom S$ for all $x\in E$. Then $\mathcal{D}^{n,S}$ is a linear mapping from $dom \mathcal{D}^{n,S}$ to space  $\mathcal{F}(X,Z)$ of all $Z$-valued functions on $X$ defined by the formula $$\mathcal{D}^{n,S}f(x)=S(f^{(n)})(x).$$
   If we choose  $X=\mathbb{R}^d$, $Y=Z=\mathbb{R}$ and $S=tr$ (trace), then the operator
 $\mathcal{D}^{2,tr}$ is Laplacian $\Delta$. If we choose  $X=Y=\mathbb{R}^d$,
  $Z=\mathbb{R}$ and $S=tr$, then the operator
 $\mathcal{D}^{1,tr}$ is divergence $div$.

 Let $E$ be a real normed space.
 Let $E$ be  continuously embedded in a separable Hilbert space $H$. Let the image of  this embedding be dense in~$H$.
Then $E\subset H\subset E^\ast$ is a rigged Hilbert space. Let $\{e_n\}$ be an orthonormal basis  in  $H$. We assume that each element of $\{e_n\}$ belongs to
$E$. The L\'evy  trace $tr_L$ 
defined by
  \begin{equation}
 \label{tlL}
 tr_L(R)=\lim_{n\to\infty}\frac 1{n}\sum_{k=1}^n(Re_k, e_k),
 \end{equation}
where  $dom\, tr_L$ consists of all $R\in L(E,E^\ast)$ for which the right side of~(\ref{tlL}) exists (see~\cite{AS1993}).
    If we choose  $X=E$, $Y=Z=\mathbb{R}$ and $S=tr_L$, then
 $\mathcal{D}^{2,tr_L}$ is the L\'evy Laplacian $\Delta_L$. If
   $X=E$, $Y=E^\ast$,
  $Z=\mathbb{R}$ and $S=tr_L$, then
 $\mathcal{D}^{1,tr_L}$ is the L\'{e}vy divergence $div_L$. An important case is then $H$ is $L_2([0,1],\mathbb{R}^d)$, $E$ is some space of curves (for example, we can choose the Cameron-Martin space of the Wiener measure as $E$), and $\{e_n\}$ is the orthonormal basis in $L_2([0,1],\mathbb{R}^d)$ defined by
$e_n(t)=p_{a_n}(t)\sqrt{2}\sin{(b_n\pi t)}$, where $\{p_1,p_2,\ldots,p_d\}$ is an orthonormal basis in $\mathbb{R}^d$, $a_n=n-4\lfloor \frac {n-1}4\rfloor$ and $b_n=\lfloor \frac {n+3}4\rfloor$ (see~~\cite{Volkov2015,Volkov2012,VolkovD}). Then the L\'evy-Laplacian associated with this basis coincides with the L\'evy-Laplacian defined in~~\cite{AGV1993,AGV1994}.\footnote{In~~\cite{AV} a 
 functional divergence has been defined as an integral functional. It is possible to show that the restriction of the L\'evy divergence associated with   basis $\{e_n\}$  to the domain of the functional divergence coincides  with that  divergence.}    The theorem about equivalence of the Yang-Mills equations and the L\'evy-Laplace equation  holds for this Laplacian.

These definitions of L\'evy differential operators can be transferred to the stochastic case.
Let  $f$ be an element of the Sobolev space $W^1_2(P)$ over the Wiener measure $P$ and   $B$ be a continuous linear mapping from the Cameron-Martin space of $P$ to $W^1_2(P)$. Let  $D^2f$ be the second derivative of $f$ and  $DB$ be the first derivative of $B$. We define the value of the stochastic L\'evy Laplacian on $f$ and the L\'evy divergence on $B$ by $tr_L D^2f$ and $tr_L DB$ respectively.
We find the value of the stochastic L\'evy Laplacian defined as the C\`esaro mean of directional derivatives on the stochastic parallel transport.
We show that, unlike the deterministic case, the equivalence of the Yang-Mills equations and the L\'evy-Laplace equation is not valid for such Laplacian. Hence, this Laplacian does not coincide with the operator introduced
in~~\cite{LV2001}.

By means of the stochastic L\'evy divergence, we study relationship between the Yang-Mills equations and infinite dimensional chiral fields. Let us recall, that the field $g\colon \mathbb{R}^d\to SU(N)$ is a general chiral field (see~~\cite{DNF}).
Its Dirihlet functional has a form
\begin{equation}
\label{chiral_field_inf}
\frac  12\int_{\mathbb{R}^d}tr(\partial_\mu g(x)\partial_\mu g^{-1}(x))dx=-\frac 12\int_{\mathbb{R}^d}tr(A_\mu (x)A^\mu(x))dx,
\end{equation}
where $A_\mu=g^{-1}(x)\partial_\mu g(x)$.
The equation of motion has a form
$$
\sum_{\mu=1}^d\partial_\mu(g^{-1}(x)\partial_\mu g(x))=0
$$
or
\begin{equation*}
\label{chiral_field}
 \left\{
\begin{aligned}
div A=0\\
\partial_\mu A_\nu-\partial_\nu A_\mu+[A_\mu,A_\nu]=0, 
\end{aligned}
\right.
\end{equation*}
where $A=(A_1,\ldots,A_d)$.
In this paper it  is proved that the connection satisfies the Yang-Mills equations if and only if the corresponding  stochastic parallel transport satisfies the analogue of the equation of motion of chiral field, where divergence is replaced with the stochastic L\'evy  divergence. Also it is proved that the Yang-Mills action functional can be represented as an analogue of~(\ref{chiral_field_inf}), where the chiral field is replaced with the stochastic parallel transport, the sum is replaced  with the C\`esaro mean and the Lebesgue measure is replaced with the product of the Lebesgue measure and the Wiener measure (cf.~~\cite{Simon}). There is no analogue of this representation in the deterministic case. On the connection between infinite dimensional analogues of chiral fields and the Yang-Mills equations in the deterministic case see~~\cite{Polyakov,AV}.\footnote{In~~\cite{AV}, in particular, an analogue of  the equation of motion of chiral field for the functional divergence has been considered. It has been shown that the deterministic parallel transport is a solution of this equation if the corresponding connection satisfies the Yang-Mills equations.}
 Note that the Dirihlet form associated with the L\'evy Laplacian has been studied in~~\cite{Volkov3AB}. Another approach  to representation of the Yang-Mills action functional  by the stochastic parallel transport has been used in~~\cite{Bauer1998}. For the recent development in the study of L\'evy Laplacians see~~\cite{AJS,AJS2015,Volkov2013}.

For the particular case of  Maxwell's equations the results of the paper have been formulated and proved in~~\cite{VolkovB1} and~~\cite{VolkovB2}.

The paper is organized as follows. In Sec. 1 we find the second derivative of the stochastic parallel transport.
In Sec. 2 we prove some technical lemmas. In Sec. 3 we define the stochastic L\'evy Laplacian as the C\`easro mean of the second-order directional derivatives and find its value on the stochastic parallel transport. In Sec. 4 we define the stochastic L\'evy divergence and obtain the equation with this divergence which is equivalent to the Yang-Mills equations. In Sec. 5 we show that  the Yang-Mills action functional can be represented as an infinitely dimensional analogue of the  Dirichlet functional of chiral fields.

\section{Stochastic parallel transport and its derivatives} 

In  this section we establish  some probabilistic  and geometric prerequisites and we find the second derivative of the stochastic parallel transport.

In the paper we use the Einstein summation convention.
Let  $(\Omega,\mathcal{F},P)$ be a canonical probability space associated with $d$-dimensional space Brownian motion on interval $[0,1]$. I.e. $$\Omega=C_0([0,1],\mathbb{R}^d):=\{\gamma\in C([0,1],\mathbb{R}^d)\colon \gamma(0)=0\},$$
$\mathcal{F}$ is the completion of the  Borel $\sigma$-field on $C_0([0,1],\mathbb{R}^d)$ with respect to the Wiener measure   $P$. We denote the $d$-dimensional Brownian motion by $b_t=(b^1_t,\ldots,b^d_t)$. We denote by  $(\mathcal{F}_t)$ the increasing family of $\sigma$-fields  generated by $b_t$.  In this paper  It\^o differentials and Stratonovich differentials are denoted  by  $db$ and by $\circ db$ respectively.

The space  $$W^{2,1}_0([0,1],\mathbb{R}^d)=\{\text{$\gamma$ is absolutely continuous, $\gamma(0)=0$, $\dot{\gamma}\in L_2((0,1),\mathbb{R}^d)$}\}$$ is the Cameron-Martin space of the Wiener measure.

Let  $\mathfrak{H}$ be a real or complex Hilbert space. By the symbol $\|\cdot\|_p$ we  denote the norm of $L_p(P,\mathfrak{H})$. The Sobolev norm $\|\cdot\|_{p,r}$ on the space $\mathcal{F}C^\infty(\mathfrak{H})$ of $\mathfrak{H}$-valued $C^\infty$-smooth cylindrical function with compact support on  $C_0([0,1],\mathbb{R}^d)$
is defined by
$$
\|f\|_{p,r}=\sum_{k=1}^r (E(\sum_{i_1\ldots i_k=1}^\infty\|\partial_{g_{i_1}}\ldots\partial_{g_{i_k}}f\|_{\mathfrak{H}}^2)^{p/2})^{1/p},
$$
 where $\{g_n\}$ is an arbitrary orthonormal basis in  $W^{2,1}_0([0,1],\mathbb{R}^d)$.
The Sobolev space
$W^p_r(P,\mathfrak{H})$ is the completion of $\mathcal{F}C^\infty(\mathfrak{H})$  with respect to the norm $\|\cdot\|_{p,r}$. 
  Here $W^\infty_r(P,\mathfrak{H})$ is the projective limit
$\projlim_{p\to +\infty}W^p_r(P,\mathfrak{H})$.

For any $h\in W^{2,1}_0([0,1],\mathbb{R}^d)$ and for any $p\geq1$ the operator $\partial_h$ can be closed as an operator from $W^p_1(P,\mathfrak{H})$  to $L_p(P,\mathfrak{H})$.
 We denote the closure of $\partial_h$ by the same symbol and by the symbol $D^{\dot{h}}$.
If $F\in W^p_r(P)$, then there exists  such $\mathfrak{H}$-valued stochastic process $D_tF$ for which  the equality 
  $D^{\dot{h}}F=\partial_h F=\int_0^1 (D_t F, \dot{h}(t))_{\mathbb{R}^d}dt$  holds for every $h\in W^{2,1}_0([0,1],\mathbb{R}^d)$.\footnote{$D_t F$ is defined almost everywhere with respect to $\lambda\times P$, where
$\lambda$ is the Lebesgue measure on $[0,1]$}
  The higher order derivatives are defined similarly (see~~\cite{Nualart}). (About different ways to define the Sobolev classes over Gaussian measures  see for example~~\cite{Volkov2Bog} and the references cited therein.)

A connection in  the trivial vector bundle with base $\mathbb{R}^d$, fiber $\mathbb{C}^N$ and structure group $U(N)$ is defined bellow as  $u(N)$-valued $C^\infty$-smooth 1-form $A(x)=A_\mu(x)dx^\mu$ on $\mathbb{R}^{d}$.
If $\phi\in C^1(\mathbb{R}^d,u(N))$, then the covariant derivative  $\phi$ in the direction of the vector field $(\frac{\partial}{\partial_\mu})$ is defined 
by the formula
$\nabla_\mu\phi=\partial_\mu\phi+[A_\mu,\phi]$. Everywhere bellow it is supposed that the connection $A$ and all its derivatives of the first and the second orders are bounded on $\mathbb{R}^d$.
The curvature corresponding to the connection $A$ is the 2-form  $F(x)=\sum_{\mu<\nu}F_{\mu\nu}(x)dx^\mu\wedge dx^\nu$, where $F_{\mu\nu}=\partial_\mu A_\nu-\partial_\nu A_\mu+[A_\mu,A_\nu]$.
Every connection $A$ satisfies the Bianchi identities
\begin{equation}
\label{Bianchi}
\nabla_{\lambda}F_{\mu\nu}+\nabla_{\nu}F_{\lambda\mu}+\nabla_{\mu}F_{\nu\lambda}=0
\end{equation}
The Yang-Mills action functional has a form
\begin{equation}
\label{YMaction}
-\int_{\mathbb{R}^d}tr (F_{\mu\nu}(x)F^{\mu\nu}(x))dx
\end{equation}
The Yang-Mills equations are  the Euler-Lagrange equations for~(\ref{YMaction}) and have a form:
\begin{equation}
\label{YMequations}
\nabla^\mu F_{\mu\nu}=0.
\end{equation}
These equations are equations on a connection $A$.

The stochastic parallel transport $U^{x}(b,t)$ ($x\in \mathbb{R}^d$) associated with the connection $A$ is a solution to the differential equation 
 (in the  sense of Stratonovich):
\begin{equation}
\label{partransp}
U^{x}(b,t)=I_N-\int_0^t A_{\mu}(x+b_s)U^{x}(b,s)\circ db^{\mu}_s
\end{equation}
Since the connection $A$ and its first-order derivatives are bounded, this equation has a unique strong solution.

We consider stochastic processes $$L^x_{\mu\nu}(b,t)=U^{x}(b,t)^{-1}F_{\mu\nu}(x+b_t)U^{x}(b,t)$$
and $$J^x_{\lambda\mu\nu}(b,t)=U^{x}(b,t)^{-1}\nabla_{\lambda}F_{\mu\nu}(x+b_t)U^{x}(b,t).$$

\begin{proposition}
\label{prop1}
The following equalities  hold:
$$
\int_0^tL^x_{\mu\nu}(b,s)u^\mu(s)\circ db^\nu_s=\int_0^tL^x_{\mu\nu}(b,s)u^\mu (s)db^\nu_s
+\frac 12\int_0^tJ_{\nu \mu}^{x\ \ \nu}(b,s)u^\mu (s)ds,
$$
\begin{equation}
\label{J}
\int_0^tJ_{\nu \mu}^{x\ \ \nu}(b,s)\circ db^\mu_s=\int_0^tJ_{\nu \mu}^{x\ \ \nu}(b,s)db^\mu_s.
\end{equation}
\end{proposition}
\begin{proof}
Indeed, since $dU^{x}(b,s)^{-1}=U^{x}(b,s)A_{\mu}(x+b_s)\circ db^{\mu}_s$ and $dU^{x}(b,s)=-A_{\mu}(x+b_s)U^{x}(b,s)\circ db^{\mu}_s$, we have
$$
dL^x_{\mu\nu}(b,s)=J^x_{\lambda\mu\nu}(b,s)\circ db^{\lambda}_s.
$$
Then\footnote{If $y_1$ and $y_2$ are semi-martingales $dy_1\cdot dy_2$ denotes the differential  of the quadratic covariation of $y_1$ and $y_2$.}
\begin{multline*}
\int_0^tL^x_{\mu\nu}(b,s)u^\mu(s)\circ db^\nu_s=\int_0^tL^x_{\mu\nu}(b,s)u^\mu(s)db^\nu_s+\frac 12\int_0^t u^\mu(s)(dL^x_{\mu\nu}(b_s))\cdot (db^\nu_s)=\\
=\int_0^tL^x_{\mu\nu}(b,s)u^\mu(s)db^\nu_s+\frac 12\int_0^tu^\mu(s) (dL^x_{\mu\nu}(b_s))\cdot (db^\nu_s)=\\
=\int_0^tL^x_{\mu\nu}(b,s)u^\mu (s)db^\nu_s
+\frac 12\int_0^tJ_{\nu \mu}^{x\ \ \nu}(b,s)u^\mu (s)ds
\end{multline*}
Similarly, we obtain
\begin{multline*}
\int_0^tJ_{\nu \mu}^{x\ \ \nu}(b,s)\circ db^\mu_s=\int_0^tJ_{\nu \mu}^{x\ \ \nu}(b,s)db^\mu_s+\frac 12\int_0^t (dJ_{\nu \mu}^{x\ \ \nu}(b,s))\cdot (db^\mu_s)=\\
=\int_0^tJ_{\nu \mu}^{x\ \ \nu}(b,s)db^\mu_s+\frac 12\int_0^t U^{x}(b,s)^{-1}\nabla_\mu\nabla_\nu F^{\mu\nu}(x+b_s)U^{x}(b,s)ds
\end{multline*}
but it is valid that $\nabla_\mu\nabla_\nu F^{\mu\nu}=0$. So, we obtain~(\ref{J}).
\end{proof}
\begin{proposition}
\label{prop2}
For any $t\in[0,1]$ it is hold that $U^x(b,t)\in W^2_\infty(P,M_N(\mathbb{C}))$.
If $u\in W^{2,1}_0([0,1],\mathbb{R}^d)$,
then
\begin{multline}
\partial_{u}U^{x}(b,t)=-U^{x}(b,t)\int_0^t L^x_{\mu\nu}(b,s)u^\mu(s)\circ db_s^\nu-A_\mu(x+b_t)u^\mu(t)U^{x}(b,t)=\\
=-U^{x}(b,t)\int_0^tL^x_{\mu\nu}(b,s)u^\mu(s) db_s^\nu-\\
-\frac 12 U^{x}(b,t)\int_0^t J_{\nu \mu}^{x\ \ \nu}(b,s)u^\mu(s)dt-A_\mu(x+b_t)u^\mu(t)U^{x}(b,t)
\end{multline}

\end{proposition}
\begin{proof}
We consider two-parameter stochastic process  $Z^{x}(b,t,s)$ defined by the formula
\begin{equation}
\label{eq1}
Z^{x}_\nu(b,t,s)=
\begin{cases}
-U^{x}(b,t)U^{x}(b,s)^{-1}A_\nu(x+b_s)U^{x}(b,s)-\\-
U^{x}(b,t)\int_s^tU^{x}(b,r)^{-1}\partial_{\nu}A_\mu(x+b_r)U^{x}(b,r)\circ db^\mu_r, &\text{if $s\leq t$}\\
0, &\text{otherwise.}
\end{cases}
\end{equation}
Let us show that $D^\nu_sU^{x}(b,t)=Z_\nu^{x}(b,t,s)$.

Theorem 2.2.1 from~~\cite{Nualart} implies that $$U^{x}(b,t)\in W^1_\infty(P,M_N(\mathbb{C}))$$ and $D^\nu_sU^{x}(b,t)$ is a solution to the equation:\footnote{The symbol $Ind_{s\leq t}$ denotes the indicator of the set $\{(s,t)\colon s\leq t\}$}
\begin{multline}
\label{eq2}
D^\nu_sU^{x}(b,t)=-A_{\nu}(x+b_s)U^{x}(b,s)Ind_{s\leq t}(s,t)-\\
-\int_0^t(D^\nu_s A_{\mu}(x+b_r)U^{x}(b,r)+ A_{\mu}(x+b_r)D^\nu_sU^{x}(b,r))\circ db^\mu_r,
\end{multline}
where
$$
D^\nu_s A_{\mu}(x+b_t)=\partial_\nu A_{\mu}(x+b_t)Ind_{s\leq t}(s,t)
$$

 We verify whether $Z^x(b,t,s)$ is the solution to equation~(\ref{eq2}).
If $s>t$, then  $Z^x(b,t,s)$ is equal to zero and obviously satisfies~(\ref{eq2}). 
 For $s\leq t$ we have
\begin{multline}
\label{eqZ3}
-A_{\nu}(x+b_s)U^{x}(b,s)
-\int_0^t(D^\nu_s A_{\mu}(x+b_r)U^{x}(b,r)+A_{\mu}(x+b_r)Z^x(b,r,s))\circ db^\mu_r=\\=-A_{\nu}(x+b_s)U^{x}(b,s)-\int_s^t(\partial_\nu A_{\mu}(x+b_r)U^{x}(b,r)-\\
-A_{\mu}(x+b_r)U^{x}(b,r)U^{x}(b,s)^{-1}A_\nu(x+b_s)U^{x}(b,s)-\\-A_{\mu}(x+b_r)\int_s^rU^{x}(b,p)^{-1}\partial_{\nu}A_\lambda(x+b_p)U^{x}(b,p)\circ db^\lambda_p)
\circ db^\mu_r=\\
=(-A_{\nu}(x+b_s)U^{x}(b,s)
+\int_s^tA_{\mu}(x+b_r)U^{x}(b,r)U^{x}(b,s)^{-1}A_\nu(x+b_s)U^{x}(b,s)\circ db^\mu_r)-\\-\int_s^t(\partial_\nu A_{\mu}(x+b_r)U^{x}(b,r)-\\-A_{\mu}(x+b_r)\int_s^rU^{x}(b,p)^{-1}\partial_{\nu}A_\lambda(x+b_p)U^{x}(b,p)\circ db^\lambda_p)\circ db^\mu_r.
\end{multline}

Note that~(\ref{partransp}) and the It\^o formula together imply
\begin{multline}
\label{ggg1}
\int_s^t(\partial_\nu A_{\mu}(x+b_r)U^{x}(b,r)-\\-A_\mu(x+b_r)U^{x}(b,r)\int_s^r(U^{x}(b,p)^{-1}\partial_{\nu}A_\lambda(x+b_p))U^{x}(b,p)\circ db^\lambda_p)
\circ db^\mu_r=\\
=\int_s^tU^{x}(b,r)(U^{x}(b,r)^{-1}\partial_\nu A_{\mu}(x+b_r)U^{x}(b,r))-\\-A_\mu(x+b_r)U^{x}(b,r)(\int_s^rU^{x}(b,p)^{-1}\partial_{\nu}A_\lambda(x+b_p)U^{x}(b,p))\circ db^\lambda_p)
\circ db^\mu_r=\\=U^{x}(b,t)\int_s^t(U^{x}(b,r)^{-1}\partial_{\nu}A_\mu(x+b_r))U^{x}(b,r)\circ db^\mu_r
\end{multline}
From~(\ref{partransp}) it follows that
$U^{x}(b,t)=U^x(b,s)-\int_s^t A_{\mu}(x+b_r)U^{x}(b,r)\circ db^{\mu}_r$. This equality  implies
\begin{multline}
\label{ggg2}
A_{\nu}(x+b_s)U^{x}(b,s)-\int_s^tA_{\mu}(x+b_r)U^{x}(b,r)U^{x}(b,s)^{-1}A_\nu(x+b_s)U^{x}(b,s)\circ db^\mu_r=\\
=(U^{x}(b,s)-\int_s^tA_{\mu}(x+b_r)U^{x}(b,r)\circ db^\mu_r)U^{x}(b,s)^{-1}A_\nu(x+b_s)U^{x}(b,s)=\\
=U^{x}(b,t)U^{x}(b,s)^{-1}A_\nu(x+b_s)U^{x}(b,s).
\end{multline}
The equalities~(\ref{ggg1}) and~(\ref{ggg2}) together imply that~(\ref{eqZ3}) is equal to
\begin{multline*}
-A_{\nu}(x+b_s)U^{x}(b,s)-\\-
U^{x}(b,t)\int_s^tU^{x}(b,r)^{-1}\partial_{\nu}A_\mu(x+b_r)U^{x}(b,r)\circ db^\mu_r=Z^x(b,t,s).
\end{multline*}
So $Z^x(b,t,s)$ coincides with $D^\nu_sU^{x}(b,t)$.

Using the It\^o formula, we finally have
\begin{multline*}
\partial_uU^{x}(b,t)=D^{\dot{u}}U^{x}(b,t)=\int_0^1Z^x_\mu(b,t,s)\dot{u}^\mu(s)ds=\\
=-\int_0^t(U^{x}(b,t)U^{x}(b,s)^{-1}A_\mu(x+b_s)U^{x}(b,s)+\\
+U^{x}(b,t)\int_s^tU^{x}(b,r)^{-1}\partial_{\mu}A_\nu(x+b_r)U^{x}(b,r)\circ db^\nu_r)\dot{u}^\mu(s)ds=\\
=-U^{x}(b,t)\int_0^tU^{x}(b,s)^{-1}F_{\mu\nu}(x+b_s)u^\mu(s)U^{x}(b,s)\circ db_s^\nu-A_\mu(x+b_t)u^\mu(t)U^{x}(b,t)=\\=-U^{x}(b,t)\int_0^t L^x_{\mu\nu}(b,s)u^\mu(s)\circ db_s^\nu-A_\mu(x+b_t)u^\mu(t)U^{x}(b,t)
\end{multline*}
Since the connection $A$ and all its derivatives of the first and the second orders are bounded, the theorem 2.2.2 from~~\cite{Nualart} implies $U^x(b,t)\in W^2_\infty(P,M_N(\mathbb{C}))$.
\end{proof}
\begin{proposition}
\label{prop3}
For any $t\in[0,1]$ it is valid that $U^x(b,t)^{-1}\in W^2_\infty(P,M_N(\mathbb{C}))$.
If $u\in W^{2,1}_0([0,1],\mathbb{R}^d)$,
then
\begin{multline}
\partial_{u}U^{x}(b,t)^{-1}=\\=(\int_0^t L^x_{\mu\nu}(b,s)u^\mu(s)\circ db_s^\nu)U^{x}(b,t)^{-1}+U^{x}(b,t)^{-1}A_\mu(x+b_t)u^\mu(t)
\end{multline}
\end{proposition}
\begin{proof}
The proof is similar to the previous one.
\end{proof}

\begin{proposition}
\label{prop4}
If $u,v\in W^{2,1}_0([0,1],\mathbb{R}^d)$ and $u(1)=v(1)=0$, then
\begin{multline}
\label{secondderivative2}
\partial_{v}\partial_{u}U^{x}(b,1)=\\
=U^{x}(b,1)\int_0^1L^x_{\mu\lambda}(b,t)u^\mu(t)(\int_0^tL^x_{\nu\kappa}(b,s)v^\nu(s)\circ db_s^\kappa)\circ db_t^\lambda+
\\+U^{x}(b,1)\int_0^1 L^x_{\nu\kappa}(b,t)v^\nu(t)(\int_0^tL^x_{\mu\lambda}(b,s)
u^\mu(s)\circ db_s^\lambda)\circ db_t^\kappa-\\
-U^{x}(b,1)(\frac 12 \int_0^1(J^x_{\nu\mu\lambda}(b,t)+J^x_{\mu\nu\lambda}(b,t))u^\mu(t)v^\nu(t)\circ db_t^\lambda+\\+\frac12 \int_0^1L^x_{\mu\nu}(b,t)(\dot{u}^\nu(t)v^\mu(t)+\dot{v}^\nu(t)u^\mu(t))dt).
\end{multline}
\end{proposition}

\begin{proof}
Lemma 1.3.4 from~~\cite{Nualart} implies
\begin{multline*}
D^\nu_s \partial_{u}U^{x}(b,1)=-Z^x_\nu(b,1,s)\int_0^1L^x_{\mu \lambda}(b,t)u^\mu(t)\circ db_t^\lambda-\\-
U^{x}(b,1)\int_0^1L^x_{\mu\lambda}(x+b_t)u^\mu(t)U^{x}(b,t)^{-1}Z^x_\nu(b,t,s)\circ db_t^\lambda-\\
+U^{x}(b,1)\int_0^1U^{x}(b,t)^{-1}Z^x_\nu(b,t,s)L^x_{\mu\lambda}(b,t)u^\mu(t)\circ db_t^\lambda-\\
-U^{x}(b,1)\int_s^1U^{x}(b,t)^{-1}\partial_\nu F_{\mu\lambda}(x+b_t)u^\mu(t)U^{x}(b,t)\circ db_t^\lambda-\\
-U^{x}(b,1)L^x_{\mu\nu}(b,s)u^\mu(s).
\end{multline*}

Since
$$
\partial_{v}\partial_{u}U^{x}(b,1)=\sum_{\nu=1}^d\int_0^1\dot{v}^\nu(s)D^\nu_s \partial_{u}U^{x}(b,1)ds,
$$
using the It\^o formula, we obtain
\begin{multline}
\label{dudvpr}
\partial_{v}\partial_{u}U^{x}(b,1)=\\
=U^{x}(b,1)\int_0^1L^x_{\mu\lambda}(b,t)u^\mu(t)(\int_0^tL^x_{\nu\kappa}(b,s)v^\nu(s)\circ db_s^\kappa)\circ db_t^\lambda+
\\+U^{x}(b,1)\int_0^1 L^x_{\nu\kappa}(b,t)v^\nu(t)(\int_0^tL^x_{\mu\lambda}(b,s)
u^\mu(s)\circ db_s^\lambda)\circ db_t^\kappa-\\
-U^{x}(b,1)(\int_0^1J^x_{\nu\mu\lambda}(b,t)u^\mu(t)v^\nu(t)\circ db_t^\lambda+\\+\int_0^1L^x_{\mu\nu}(b,t)\dot{v}^\nu(t)u^\mu(t)dt).
\end{multline}

The It\^o formula implies:
\begin{multline}
\label{dudvpr2}
-\frac 12U^{x}(b,1)\int_0^1L^x_{\mu\nu}(b,s)u^\mu(s)\dot{v}^\nu(s)ds=
\\=\frac 12 U^{x}(b,1)\int_0^1J^x_{\lambda\mu\nu}(b,s)u^\mu(s)v^\nu(s)\circ db^\lambda_s+\\
+\frac 12U^{x}(b,1)\int_0^1L^x_{\mu\nu}(b,s)\dot{u}^\mu(s)v^\nu(s)ds=\\
=-\frac 12 U^{x}(b,1)\int_0^1J^x_{\mu\nu\lambda}(b,s)u^\mu(s)v^\nu(s)\circ db^\lambda_s-\\
-\frac 12 U^{x}(b,1)\int_0^1J^x_{\nu\lambda\mu}(b,s)u^\mu(s)v^\nu(s)\circ db^\lambda_s+\\
+\frac 12U^{x}(b,1)\int_0^1L^x_{\mu\nu}(b,s)\dot{u}^\mu(s)v^\nu(s)ds.
\end{multline}
The last equality holds due to the Bianchi identities~(\ref{Bianchi}).
Then equalities~(\ref{dudvpr}) and~(\ref{dudvpr2}) together imply~(\ref{secondderivative2}).
\end{proof}

\section{Technical lemmas}

There are certain technical results obtained in this section that are referred to in proofs of the theorems in the following sections.

Let $h_n(t)=\sqrt{2}\sin{n\pi t}$  and
$
l_n(s,t)=\frac 1n \sum_{k=1}^n h_k(s)h_k(t)$.
Note that
\begin{equation}
\label{sin}
\lim_{n\to\infty}l_n(s,t)=\begin{cases}
1, &\text{if $t=s$ and $(t,s)\neq(0,0),(1,1)$}\\
0, &\text{otherwise}
\end{cases}
\end{equation}
and for all $n\in \mathbb{N}$ and $(t,s)\in [0,1]\times[0,1]$ the following inequality holds:
\begin{equation}
\label{sin2}
|l_n(s,t)|\leq2.
\end{equation}

Bellow $\|\cdot\|$ denotes the Frobenius norm on the space $M_N(\mathbb{C})$.

\begin{lemma}
\label{lemmadbdb}
Let $H_\mu, M_\mu$ be adapted with $(\mathcal{F}_t)$  bounded $M_N(\mathbb{C})$-valued stochastic processes.
Then in $L_2(P,M_N(\mathbb{C}))$ the following equalities hold
\begin{equation}
\label{l11}
\lim_{n\to\infty}\int_0^1(\int_0^t H_\mu(b,s)l_n(s,t)db_s^\mu) M_\nu(b,t)db^\nu_t=0,
\end{equation}
\begin{equation}
\label{l1}
\lim_{n\to\infty}\int_0^1 M_\nu(b,t)(\int_0^t H_\mu(b,s)l_n(s,t)db_s^\mu) db^\nu_t=0.
\end{equation}
\end{lemma}
\begin{proof}
In the proof of the lemma we denote $$\int_0^1(\int_0^t H_\mu(b,s)l_n(s,t)db_s^\mu) M_\nu(b,t)db^\nu_t$$ by
$R_n(b)$. 
Due to the Fubini--Tonelli theorem we have
\begin{multline*}
\|R_n\|_2^2=\\=E(tr((\int_0^1(\int_0^t H_\mu(b,s)l_n(s,t)db_s^\mu)M_\nu(b,t)db_t^\nu)\times\\\times(\int_0^1(\int_0^t H_\mu(b,s)l_n(s,t)db_s^\mu) M_\nu(b,t)db_t^\nu)^\ast))=\\
=\int_0^1E(tr((\int_0^t H^\ast_\mu(b,s)l_n(s,t)db_s^\mu)(\int_0^t H_\mu(b,s)l_n(s,t)db_s^\mu) M^\nu(b,t)M_\nu^\ast(b,t)))dt
\end{multline*}
Since the processes $M_\mu$ are bounded, there exists such a constant $C>0$ that
\begin{multline*}
\|R_n\|_2^2\leq C\int_0^1E\|\int_0^t H_\mu(b,s)l_n(s,t)db_s^\mu\|^2dt=\\=C\int_0^1\int_0^tE(tr(H^\ast_\mu(b,s)H^\mu(b,s)))l^2_n(s,t)dsdt. 
\end{multline*}
Then  equality~(\ref{l11}) follows from~(\ref{sin}),~(\ref{sin2}), boundedness of processes $H_\mu$ and Lebesgue's dominated convergence 
theorem.
Equality~(\ref{l1}) can be proved similarly.

\end{proof}

\begin{lemma}
\label{lemmadtdb}
Let $M_\mu, K$ be adapted with $(\mathcal{F}_t)$ bounded $M_N(\mathbb{C})$-valued processes.  
Then in $L_2(P,M_N(\mathbb{C}))$ the following equalities hold
\begin{equation}
\label{l22}
\lim_{n\to\infty}\int_0^1(\int_0^t K(b,s)l_n(s,t)ds) M_\nu(b,t)db^\nu_t=0,
\end{equation}
\begin{equation}
\label{ldtdb2}
\lim_{n\to\infty}\int_0^1 M_\nu(b,t)(\int_0^t K(b,s)l_n(s,t)ds) db^\nu_t=0.
\end{equation}
\end{lemma}
\begin{proof}
In the proof of the lemma we denote
$\int_0^1\int_0^t K(b,s)l_n(s,t)ds M_\nu(b,t)db^\nu_t$ by
$R_n(b)$. 
Due to the Fubini--Tonelli theorem we have
\begin{multline*}
\|R_n\|_2^2=E(tr(\int_0^1(\int_0^t K(b,s)l_n(s,t)ds) M_\nu(b,t)db_t^\nu)\times\\\times(\int_0^1(\int_0^t K(b,s)l_n(s,t)ds) M_\nu(b,t)db_t^\nu)^\ast))=\\
=E(tr(\int_0^1(\int_0^t K^\ast(b,s)l_n(s,t)ds)(\int_0^t K(b,s)l_n(s,t)ds) M^\nu(b,t)M_\nu^\ast(b,t)dt))=\\
=\int_0^1\int_0^t\int_0^tE( tr(K^\ast(b,s_1)K(b,s_2)M^\nu(b,t)M_\nu^\ast(b,t)))l_n(s_1,t)l_n(s_2,t)ds_1ds_2dt
\end{multline*}
Since the processes  $M_\mu, K$ are bounded, there exists such a constant $C>0$ that
$$
\|R_n\|_2^2\leq C\int_0^1\int_0^t\int_0^t |l_n(s_1,t)l_n(s_2,t)|ds_1ds_2dt.
$$
Then equality~(\ref{l22}) follows from~(\ref{sin}) and~(\ref{sin2}) and Lebesgue's dominated convergence 
theorem.
Equality~(\ref{ldtdb2}) can be proved similarly.
\end{proof}

\begin{lemma}
\label{lemmadbdt}
Let $H_\mu$, $K$ be adapted with $(\mathcal{F}_t)$   bounded $M_N(\mathbb{C})$-valued processes.  
Then in  $L_2(P,M_N(\mathbb{C}))$ the following equalities hold
\begin{equation}
\label{l31}
\lim_{n\to\infty}\int_0^1(\int_0^t H_\mu(b,s)l_n(s,t)db_s^\mu) K(b,t)dt=0,
\end{equation}
\begin{equation}
\label{l32}
\lim_{n\to\infty}\int_0^1 K(b,t)(\int_0^t H_\mu(b,s)l_n(s,t)db_s^\mu) dt=0.
\end{equation}
\end{lemma}
\begin{proof}
In the proof of the lemma we denote $\int_0^1(\int_0^t K(b,s)l_n(s,t)ds) M_\nu(b,t)db^\nu_t$ by
$R_n(b)$.
Due to the Fubini--Tonelli theorem we have
\begin{multline*}
\|R_n\|_2^2=\\
=\int_0^1\int_0^1E(tr((\int_0^{t_1} H^\ast_\mu(b,s_1)l_n(s_1,t_1)db_{s_1}^\mu)\times\\\times(\int_0^{t_2} H_\mu(b,{s_2})l_n(s_2,t_2)db_{s_2}^\mu) K(b,t_2)K^\ast(b,t_1)))dt_1dt_2
\end{multline*}
Since the process  $K$ is bounded, there exists such a constant $C>0$ that
\begin{multline*}
\|R_n\|_2^2\leq \\
\leq C \int_0^1\int_0^1E\|\int_0^{t_1} H_\mu(b,s_1)l_n(s_1,t_1)db_{s_1}^\mu\|\times\\ \times\|\int_0^{t_2} H_\mu(b,{s_2})l_n(s_2,t_2)db_{s_2}^\mu\|dt_1dt_2.
\end{multline*}
Then
\begin{multline*}
\|R_n\|_2^2\leq \\
\leq \frac 12C E(\int_0^1\int_0^1\|\int_0^{t_1} H_\mu(b,s_1)l_n(s_1,t_1)db_{s_1}^\mu)\|\times\\\times
\|\int_0^{t_2} H_\mu(b,{s_2})l_n(s_2,t_2)db_{s_2}^\mu)\|dt_1dt_2)\leq\\ \leq
 \frac 12 C \int_0^1\int_0^1E(\|\int_0^{t_1} H_\mu(b,s_1)l_n(s_1,t_1)db_{s_1}^\mu\|^2+\\+\|\int_0^{t_2} H_\mu(b,{s_2})l_n(s_2,t_2)db_{s_2}^\mu\|^2)dt_1dt_2=
\\=C \int_0^1E(\|\int_0^{t} H_\mu(b,s)l_n(s,t)db_{s}^\mu\|^2)dt=\\
= C\int_0^1\int_0^{t}  E(tr(H^\mu(b,s) H^\ast_\mu(b,s))l^2_n(s,t)dsdt
\end{multline*}
Then equality~(\ref{l31}) follows from~(\ref{sin}) and~(\ref{sin2}) and Lebesgue's dominated convergence 
theorem.
Equality~(\ref{l32}) can be proved similarly.
\end{proof}
\begin{lemma}
\label{lemmadtdt}
Let $R, K$ be adapted with $(\mathcal{F}_t)$ bounded $M_N(\mathbb{C})$-valued processes.  
Then in $L_2(P,M_N(\mathbb{C}))$ the following equalities hold
\begin{equation}
\label{lemmadtdt1}
\lim_{n\to\infty}\int_0^1(\int_0^t R(b,s)l_n(s,t)ds) K(b,t)dt=0,
\end{equation}
\begin{equation}
\label{lemmadtdt2}
\lim_{n\to\infty}\int_0^1 K(b,t)(\int_0^t R(b,s)l_n(s,t)ds)dt=0.
\end{equation}
\end{lemma}
\begin{proof}
Denote $\int_0^1(\int_0^t R(b,s)l_n(s,t)ds) K(b,t)dt$  by
$R_n(b)$.
Due to the Fubini--Tonelli theorem we have
\begin{multline*}
\|R_n\|_2^2=\int_0^1\int_0^1\int_0^{t_1}\int_0^{t_2}l_n(s_1,t_1)l_n(s_2,t_2)\times\\\times E(tr(R^\ast(b,s_1)R(b,s_2)K(b,t_1)K^\ast(b,t_2)))ds_1ds_2dt_1dt_2
\end{multline*}
Then equality~(\ref{lemmadtdt1}) follows from~(\ref{sin}),~(\ref{sin2}), boundedness of processes $R, K$  and Lebesgue's dominated convergence 
theorem.
Equality~(\ref{lemmadtdt2}) can be proved similarly. 
\end{proof}

\section{Stochastic L\'evy Laplacian}

\begin{definition}
The stochastic L\'evy Laplacian  $\Delta_L$ is a  linear mapping from $dom \Delta_L$ to $L_2(P,M_N(\mathbb{C}))$ defined as
\begin{equation}
\label{LaplaceL}
\Delta_Lf(b)=\lim_{n\to\infty}\frac 1n \sum_{k=1}^n \sum_{\mu=1}^d\partial_{p_\mu h_k}\partial_{p_\mu h_k}f(b), 
\end{equation}
where the sequence converges in $L_2(P,M_N(\mathbb{C}))$ and 
$dom \Delta_L$  consists of all $f\in W^2_2(P,M_N(\mathbb{C}))$ for which the right-hand side of~(\ref{LaplaceL})
exists.
\end{definition}

\begin{remark}
\label{rk1}
If $f\in dom \Delta_L$,  then $\Delta_Lf(b)=d\lim_{n\to\infty}\frac 1n \sum_{k=1}^n\partial_{e_n}\partial_{e_n}f(b)$, where, as it has been mentioned in the introduction,  $\{e_n\}$ is the orthonormal basis in $L_2([0,1],\mathbb{R}^d)$ defined by
$e_n(t)=p_{a_n}(t)h_{b_n}(t)$, where
 $a_n=n-4\lfloor \frac {n-1}4\rfloor$ and $b_n=\lfloor \frac {n+3}4\rfloor$. This formula precises the
 definition to  $tr_L D^2f$.
 The stochastic analogue of the L\'evy d'Alambertian (see~~\cite{AGV1993,AGV1994,VolkovD,VolkovB1}) can be defined as 
 \begin{equation}
\label{d'Alambertian}
\Box_Lf(b)=\lim_{n\to\infty}\frac 1n \sum_{k=1}^n \partial_{p_1 h_k}\partial_{p_1 h_k}f(b)-\sum_{\mu=2}^d\lim_{n\to\infty}\frac 1n \sum_{k=1}^n \partial_{p_\mu h_k}\partial_{p_\mu h_k}f(b).
\end{equation}
\end{remark}

\begin{theorem}
\label{thm1}
The following equality holds
$$
\Delta_LU^{x}(b,1)=U^{x}(b,1)(\int_0^1L^x_{\mu\nu}(b,t)L^{x\mu\nu}(b,t)dt-
\int_0^1J^{x\mu}_{\ \ \mu \nu}(b,s)\circ db^\nu_t)
$$
\end{theorem}
\begin{proof}
Introduce the notations
$$
R_n^{x}(b)=\frac 1n
\sum_{k=1}^n\sum_{\mu=1}^d\partial_{p_\mu h_k}\partial_{p_\mu h_k}U^{x}(b,1)
$$
and
$$
R^{x}(b)=U^{x}(b,1)(\int_0^1L^x_{\mu\nu}(b,t)L^{x\mu\nu}(b,t)dt-\int_0^1J^{x\mu}_{\ \ \mu \nu}(b,t)\circ db^\nu_t)
$$
Proposition~\ref{prop4} implies
\begin{multline*}
\label{RRx}
\|R^{x}-R_n^{x}\|_2^2
=-E(tr(\int_0^1L^x_{\mu\nu}(b,t)L^{x\mu\nu}(b,t)(1-l_n(t,t))dt+\\
+\int_0^1(\int_0^tL^x_{\mu\lambda}(b,s)l_n(s,t)\circ db_s^\lambda) L^{x\mu}_{\ \ \nu}(b,t)db_t^\nu+\\
+\frac 12\int_0^1(\int_0^tL^x_{\mu\lambda}(b,s)l_n(s,t)\circ db_s^\lambda) J_{\nu}^{x\ \mu\nu}(b,t)dt+\\
+\int_0^1J^{x\mu}_{\ \ \mu \nu}(b,t)(1-l_n(t,t))\circ db^\nu_t)^2).
\end{multline*}
It  follows from lemmas~\ref{lemmadbdb} and~\ref{lemmadtdb} that
\begin{equation}
\label{thmeq1}
-\lim_{n\to\infty}E(tr(\int_0^1(\int_0^tL^x_{\mu\lambda}(b,s)l_n(s,t)\circ db_s^\lambda) L^{x\mu}_{\ \ \nu}(b,t)db_t^\nu)^2)=0
\end{equation}
It  follows from lemmas~\ref{lemmadbdt} and~\ref{lemmadtdt} that
\begin{equation}
\label{thmeq2}
-\lim_{n\to\infty}E(tr(\int_0^1(\int_0^tL^x_{\mu\lambda}(b,s)l_n(s,t)\circ db_s^\lambda)J_{\nu}^{x\ \mu\nu}(b,t)dt)^2)=0
\end{equation}
Lebesgue's dominated convergence 
theorem implies
\begin{equation}
\label{thmeq3}
\lim_{n\to\infty}E(tr(\int_0^1L^x_{\mu\nu}(b,t)L^{x\mu\nu}(b,t)(1-l_n(t,t))dt)^2)=0
\end{equation}
and
\begin{equation}
\label{thmeq4}
\lim_{n\to\infty}E(tr(\int_0^1J_{\mu \nu}^{x\ \ \mu}(b,t)(1-l_n(t,t))\circ db^\nu_t)^2)=0.
\end{equation}
Then  Minkowski inequality and~(\ref{thmeq1}),~(\ref{thmeq2}),~(\ref{thmeq3}),~(\ref{thmeq4}) together imply that $$\lim_{n\to\infty}\|R^{x}-R_n^{x}\|_2=0.$$
\end{proof}

\begin{remark}
If  the connection $A$ satisfies the Yang-Mills equations~(\ref{YMequations}), it is valid that  $\Delta_LU^{x}(b,1)=U^{x}(b,1)\int_0^1L^x_{\mu\nu}(b,t)L^{x\mu\nu}(b,t)dt$.
It would be interesting to study relationship between the Laplacians introduced in the present paper and in
~~\cite{LV2001}.
\end{remark}
\begin{remark}
There exists the canonical unitary 
isomorphism between   $L_2(P)$ and $\Gamma(L_2([0,1],\mathbb{R}^d))$ (the boson Fock space over the Hilbert space $L_2([0,1],\mathbb{R}^d)$). In the white noise theory a rigged Hilbert space  $\mathcal E\subset \Gamma(L_2([0,1],\mathbb{R}^d))\subset \mathcal E^\ast$ is  considered, there $\mathcal E$ is the space of white noise test functionals and
$\mathcal E^\ast$ is the space of white noise generalized functionals. The L\'evy Laplacian on the space of white noise generalized functionals  has been defined by  T. Hida in~~\cite{H1975}
(see also~~\cite{HidaSi,K,KOS}). This operator is equal to zero on $\Gamma(L_2([0,1],\mathbb{R}^d))$ (see~~\cite{K}). Hence, it does not coincides with the L\'evy Laplacian introduced in the present paper. But in the next paper we will show that the last operator  coincides with  an element from the chain of nonclassical L\'evy Laplacians acting on the space of white noise generalized functionals (see~~\cite{AS007,Volkov2013}).
Moreover, we assume that this operator can be represented as $\int_{|s-t|<\varepsilon}\dot{a}_\mu (s)\dot{a}^\mu(t) dsdt$, where $\dot{a}^\mu(t)$ is the derivative of the annihilation process.
\end{remark}

\section{Stochastic L\'evy Divergence}

\begin{definition} The stochastic L\'evy divergence $div_L$ is a linear mapping from  $dom\,div_L$ to $L_2(P,M_N(\mathbb{C}))$ defined by
the formula
\begin{equation}
\label{DivL}
div_LB(b)=\lim_{n\to\infty}\frac 1n\sum_{k=1}^n \sum_{\mu=1}^d\partial_{p_\mu h_k}B(b)(p_\mu h_k), 
\end{equation}
where the sequence converges in $L_2(P,M_N(\mathbb{C}))$ and 
$dom\,div_L$  consists of all $$B\in L(W^{2,1}_0([0,1],\mathbb{R}^d),W^1_{2}(P,M_N(\mathbb{C})))$$ for which the right-hand side  of~(\ref{DivL})
exists.
\end{definition}
\begin{remark}
We use  the notations of remark~\ref{rk1}. If $B\in dom\,div_L$,  then 
$$div_LB(b)=d\lim_{n\to\infty}\frac 1n \sum_{k=1}^n\partial_{e_n}B(b)e_n.$$
This formula precises the definition to  $tr_L DB$.
\end{remark}

\begin{example}
Let $d=3$. Let $\epsilon_{\mu\nu\lambda}$ be totally antisymmetric unit tensor. 
Let
$$
S^x(b)h=\int_0^1 \epsilon_{\mu\nu\lambda}L^{x\nu\lambda}(b,t)h^\mu(t)dt
$$
The Bianchi  identities
imply that
$$
div_L S^x(b)=\int_0^1 \epsilon_{\mu\nu\lambda}J^{x\mu\nu\lambda}(b,t)dt=0.
$$
\end{example}

From the fact that $U^x(b,1),U^x(b,1)^{-1} \in  W^2_\infty(P,M_N(\mathbb{C}))$ it follows that $$B^{A,x}\in L(W^{2,1}_0([0,1],\mathbb{R}^d),W^1_{2}(P,M_N(\mathbb{C}))),$$ where $B^{A,x}$
is  defined by the formula
\begin{multline*}
B^{A,x}(b)u=U^{x}(b,1)^{-1}\partial_uU^{x}(b,1)=\\
=-\int_0^1U^{x}(b,t)^{-1}F_{\mu\nu}(x+b_t)u^\mu(t)U^{x}(b,t)\circ db_t^\nu-\\
U^{x}(b,1)^{-1}A_\mu(x+b_1)u^\mu(1)U^{x}(b,1).
\end{multline*}
Note that in fact  $B^{A,x}\in L(W^{2,1}_0([0,1],\mathbb{R}^d),W^1_{2}(P,u(N))$.

\begin{proposition}
\label{secf}
If $u,v\in W^{2,1}_0([0,1],\mathbb{R}^d)$ and $u(1)=v(1)=0$, then
\begin{multline}
\label{derivativeB}
\partial_uB^{A,x}(b)v
=
\int_0^1 [L^x_{\nu\kappa}(b,t)v^\nu(t),\int_0^tL^x_{\mu\lambda}(b,s)
u^\mu(s)\circ db_s^\lambda]\circ db_t^\kappa-\\
-\frac 12 \int_0^1(J^x_{\nu\mu\lambda}(b,t)+J^x_{\mu\nu\lambda}(b,t))u^\mu(t)v^\nu(t)\circ db_t^\lambda-\\-\frac12 \int_0^1L^x_{\mu\nu}(b,t)(\dot{u}^\nu(t)v^\mu(t)+\dot{v}^\nu(t)u^\mu(t))dt.
\end{multline}

\end{proposition}

\begin{proof}
It is valid that
\begin{equation}
\partial_u B^{A,x}(b,v)=\partial_v U^x(b,1)^{-1}\partial_u U^x(b,1)+U^x(b,1)^{-1}\partial_v \partial_u U^x(b,1).
\end{equation}
This equality and  Propositions~\ref{prop2},~\ref{prop3} and~\ref{prop4}  together imply~(\ref{derivativeB}).
\end{proof}

\begin{theorem}
The following equality holds
\begin{multline}
\label{divB}
div_L B^x(b)=-\int_0^1J^{x\mu}_{\ \ \mu \nu}(b,s)\circ db^\nu_s=\\
=-\int_0^1U^{x}(b,s)^{-1}\nabla^\mu F_{\mu\nu}(x+b_s)U^{x}(b,s)db_s^\nu.
\end{multline}
\end{theorem}

\begin{proof}
The proof is similar to the proof of Theorem~\ref{thm1}.
Due to Proposition~\ref{secf}
we have
\begin{multline}
\label{derivativeB1}
\sum_{\mu=1}^d\partial_{p_\mu h_n}B^{A,x}(b)(p_\mu h_n)=\int_0^1[L^x_{\mu\lambda}(b,t)h_n(t),\int_0^tL^{x\mu}_{\ \ \kappa}(b,s)h_n(s)\circ db_s^\kappa]\circ db_t^\lambda-
\\
- \int_0^1J^{x\mu}_{\ \ \mu\nu}(b,t)h_n^2(t)\circ db_t^\nu=\\
=\int_0^1[L^{x}_{\mu\lambda}(b,t), \int_0^tL^{x\mu}_{\ \ \kappa}(b,s)l_n(s,t)\circ db_s^\kappa]db_t^\lambda+\\
+\frac 12\int_0^1[ J_{\lambda \mu}^{x\ \ \lambda}(b,t),\int_0^tL^{x\mu}_{\ \ \kappa}(b,s)l_n(s,t)\circ db_s^\kappa]dt-\\
-\int_0^1J^{x\mu}_{\ \ \mu\nu}(b,t)h_n^2(t)\circ db_t^\nu.
\end{multline}
Then Lemmas~\ref{lemmadbdb},~\ref{lemmadbdt},~\ref{lemmadtdb},~\ref{lemmadtdt},  Lebesgue's dominated convergence theorem  and the Minkowski inequality  imply~(\ref{divB}).
\end{proof}

\begin{theorem}
The following two assertions are equivalent:
\begin{enumerate}
\item  a connection $A$ satisfies the Yang-Mills equations: $\nabla_\mu F^{\mu}_{\ \nu}=0$,
\item  $div_LB^{A,x}=0$ for some $x\in\mathbb{R}^d$.
\end{enumerate}
\end{theorem}
\begin{proof}
Let it is valid for some $x\in\mathbb{R}^d$  that $div_LB^{A,x}=0$.
Then
\begin{multline*}
0=-E(tr(div_LB^{A,x})^2)=-E\int_0^1 tr(\nabla_\mu F^{\mu}_{\ \nu}(x+b_t)\nabla_\lambda F^{\lambda\nu}(x+b_t))dt=\\
=-\int_0^1\int_{\mathbb{R}^d}tr(\nabla_\mu F^{\mu}_{\ \nu}(x+y)\nabla_\lambda F^{\lambda\nu}(x+y)){(2\pi t)}^{-\frac d2}e^{-\frac {(y,y)_{\mathbb{R}^d}^2}{2t}}dydt
\end{multline*}
Then we obtain $\nabla_\mu F^{\mu}_{\ \nu}=0$.
The other side of the proof of the theorem is trivial.
\end{proof}

\section{Action functional}

In this section it is proved that the Yang-Mills action functional~(\ref{YMaction}) can be represented as  an infinite-dimensional analogue of  the Dirichlet functional of chiral field.

\begin{proposition}
\label{prop6}
The following equality holds
\begin{multline*}
\lim_{n\to\infty}\frac 1n \sum_{k=1}^n E(tr(\partial_{p_\mu h_k}U^{x}(b,1)^{-1}\partial_{p_\mu h_k}U^{x}(b,1)))=\\=-E(\int_0^1 tr(F_{\mu\nu}(x+b_t)F_{\mu}^{\ \nu}(x+b_t))dt).
\end{multline*}
\end{proposition}
\begin{proof}
Note that Propositions~\ref{prop2} and~\ref{prop3} imply
$$
tr(\partial_{p^\mu h_n}U^{x}(b,1)^{-1}\partial_{p^\mu h_n}U^{x}(b,1))=-tr(\int_0^1L^x_{\mu\nu}(b,t)h_n(t)\circ db^\nu_t)^2.
$$
Using the It\^o formula we have
\begin{multline}
(\int_0^1L^x_{\mu\nu}(b,t)h_n(t)\circ db^\nu_t)^2=\int_0^1L^x_{\mu\nu}(b,t)L_{\mu}^{x \ \nu}(b,t)h^2_n(t)dt+\\
+2\int_0^1 (\int_0^tL^x_{\mu\nu}(b,s)h_n(s)\circ db^\nu_s) L^x_{\mu\lambda}(b,t)h_n(t)db_t^\lambda+\\+
\int_0^1(\int_0^tL^x_{\mu\nu}(x+b_s)h_n(s)\circ db^\nu_s)J_{\lambda \mu}^{x\ \ \lambda}(b,t)h_n(t)dt.
\end{multline}
It is valid that
$$
E(\int_0^1 (\int_0^tL^x_{\mu\nu}(b,s)h_n(s)\circ db^\nu_s) L^x_{\mu\lambda}(b,t)h_n(t)db_t^\lambda)=0.
$$
Lemmas~\ref{lemmadbdt} and~\ref{lemmadtdt} imply
$$
\lim_{n\to\infty}E(\int_0^1(\int_0^tL^x_{\mu\nu}(b,s)l_n(s,t)\circ db^\nu_s)J_{\lambda \mu}^{x\ \ \lambda}(b,t)dt)=0.
$$
We obtain
\begin{multline}
\lim_{n\to\infty}\frac 1n \sum_{k=1}^n  E(tr(\partial_{p_\mu h_k}U^{x}(b,1)^{-1}\partial_{p_\mu h_k}U^{x}(b,1)))=\\
=-\lim_{n\to\infty}E(tr\int_0^1L^x_{\mu\nu}(b,t)L_{\mu}^{x\ \nu}(b,t)l_n(t,t)dt)=\\=-E(\int_0^1 tr(F_{\mu\nu}(x+b_t)F_{\mu}^{\ \nu}(x+b_t))dt).
\end{multline}
The last equality follows from Lebesgue's dominated convergence 
theorem.
\end{proof}

\begin{theorem}
If
$$
-\int_{\mathbb{R}^d}tr(F_{\mu\nu}(x)F^{\mu\nu}(x))dx<\infty
$$
and
$$
-\int_{\mathbb{R}^d}tr(\nabla_\mu F^{\mu}_{\ \nu}(x)\nabla_\lambda F^{\lambda\nu}(x))dx<\infty ,
$$
then
\begin{multline}
\label{E1}
\lim_{n\to\infty}\frac 1n\sum_{k=1}^n \sum_{\mu=1}^d \int_{\mathbb{R}^d}E(tr(\partial_{p_\mu h_k}U^{x}(b,1)^{-1}\partial_{p_\mu h_k}U^{x}(b,1))dx)=\\=-\int_{\mathbb{R}^d}tr(F_{\mu\nu}(x)F^{\mu\nu}(x))dx
\end{multline}
\end{theorem}
\begin{proof}
Introduce the notations
$$
G_1(x)=-E(\int_0^1tr(F_{\mu\nu}(x+b_t)F^{\mu\nu}(x+b_t))dt),
$$
$$
G_2(x)=-E(\int_0^1tr(\nabla_\mu F^{\mu}_{\ \lambda}(x+b_t)\nabla_\nu F^{\nu\lambda}(x+b_t))dt).
$$
The Fubini--Tonelli theorem implies
\begin{multline*}
\int_{\mathbb{R}^d}G_1(x)dx=-E(\int_0^1\int_{\mathbb{R}^d}tr(F_{\mu\nu}(x+b_t)F^{\mu\nu}(x+b_t))dxdt)=\\
=-\int_{\mathbb{R}^d}tr(F_{\mu\nu}(x)F^{\mu\nu}(x))dx<\infty,
\end{multline*}
\begin{multline*}
\int_{\mathbb{R}^d}G_2(x)dx=-E(\int_0^1\int_{\mathbb{R}^d}tr(\nabla_\mu F^{\mu}_{\ \lambda}(x)\nabla_\nu F^{\nu\lambda}(x))dxdt)=\\=-\int_{\mathbb{R}^d}tr(\nabla_\mu F^{\mu}_{\ \lambda}(x)\nabla_\nu F^{\nu\lambda}(x))dx<\infty.
\end{multline*}
Note that
\begin{multline*}
- E(tr(\int_0^1L^x_{\mu\nu}(b,t)h_k(t)\circ db^\nu_t)^2)
\leq -2E(tr(\int_0^1L^x_{\mu\nu}(x+b_t)h_k(t)db^\nu_t)^2)-\\
-\frac 12E(tr(\int_0^1J_{\nu \mu}^{x\ \ \nu}(b,t)h_k(t)dt)^2)
\end{multline*}
Since~(\ref{sin2}) holds, for all $n\in\mathbb{N}$ we have
$$
-\frac 1n\sum_{k=1}^n\sum_{\mu=1}^dE(tr(\int_0^1L^x_{\mu\nu}(x+b_t)h_k(t)db^\nu_t)^2)\leq 
2G_1(x)
$$
and
\begin{multline*}
-\frac 1n\sum_{k=1}^n\sum_{\mu=1}^dE(tr(\int_0^1J_{\nu \mu}^{x\ \ \nu}(b,t)h_k(t)dt)^2)=\\
=-\frac 1n\sum_{k=1}^n E(tr(\int_0^1\int_0^1J_{\nu \mu}^{x\ \ \nu}(b,s)J_{\lambda}^{x\ \mu \lambda}(b,t)h_k(t)
h_k(s)dtds))\leq\\
\leq -\frac 1n\sum_{k=1}^n\frac12\int_0^1\int_0^1E(tr(J_{\nu \mu}^{x\ \ \nu}(b,t)J_{\lambda}^{x\ \mu\lambda}(b,t))h_k^2(t)+\\+
tr(J_{\nu \mu}^{x\ \ \nu}(b,s)J_{\lambda}^{x\ \mu\lambda}(b,s))h_k^2(s))dtds
\leq 2G_2(x)
\end{multline*}

Then equality~(\ref{E1}) follows from  Proposition~\ref{prop6} and Lebesgue's dominated convergence 
theorem.

\end{proof}

\section*{Acknowledgments}

The author would like to express his deep gratitude to  O.~G.~Smolyanov and I.~V.~Volovich for many helpful discussions.

This work is supported by the Russian Science Foundation under grant 14-50-00005.

\end{document}